\documentclass[journal]{IEEEtran}

\usepackage{bm}
\ifCLASSINFOpdf
\else
   \usepackage[dvips]{graphicx}
\fi
\usepackage{graphicx}
\usepackage{hyperref}
\usepackage{cite}
\usepackage{amsmath, amsfonts}
\usepackage{amssymb}
\usepackage{booktabs} % 用于产生更专业的表格横线
\usepackage{multirow} % 用于合并多行
\usepackage{caption}
\usepackage{subcaption}
\usepackage{upgreek}
\usepackage{pifont}
\usepackage{xcolor}

\begin{document}

\title{CoFi-Lite: Pushing the Limits of Ultra-Lightweight Speech Enhancement}

\author{Leyan Yang, Dahan Wang, Xiaobin Rong, Jiadong Zhao, Jing Lu, 
\IEEEmembership{Senior Member, IEEE}
\thanks{
This work was supported by the National Natural Science Foundation of China with grant no. 12274221. (Corresponding author: Jing Lu.)}
\thanks{The authors are with the Key Laboratory of Modern Acoustics, Nanjing University, Nanjing 210093, China, and also with NJU-Horizon Intelligent Audio Lab, Horizon Robotics, Beijing 100094, China (e-mail: leyan.yang@smail.nju.edu.cn; dahan.wang@smail.nju.edu.cn; xiaobin.rong@smail.nju.edu.cn; jiadong.zhao@smail.nju.edu.cn; lujing@nju.edu.cn).}}

\maketitle

\begin{abstract}
Ultra-lightweight models are essential for the deployment of deep learning-based speech enhancement algorithms on edge devices.
Although recent approaches have achieved a certain balance between computational complexity and performance, pushing the complexity limits further demands more sophisticated designs.
In this letter, we propose CoFi-Lite, a highly efficient model that decouples spectral modeling into coarse- and fine-grained streams.
By leveraging two parallel and symmetric encoder-decoder paths, it simultaneously extracts full-band envelopes and low-frequency details for complementary enhancement.
In addition, a novel Cross-Path Fusion (CPF) module is introduced to bridge the distinct paths, facilitating efficient feature interaction.
Remarkably, CoFi-Lite requires extremely low computational resources, featuring only 12.87M MACs/s and 83.12k parameters.
Experimental results demonstrate that our proposed model outperforms the ultra-lightweight baseline GTCRN while requiring only 40.26\% of its computational complexity. 
Its scaled-up variant also delivers performance on par with that of the SOTA ultra-lightweight model AdaptCRN alongside a 19.34\% reduction in computational cost.
Audio examples are available at \href{https://acceleration123.github.io/CoFiLite-demo/}{https://acceleration123.github.io/CoFiLite-demo/}.
\end{abstract}

\begin{IEEEkeywords}
speech enhancement,
ultra-lightweight model,
computational complexity
\end{IEEEkeywords}

\IEEEpeerreviewmaketitle

\section{Introduction}
\IEEEPARstart{S}{peech} enhancement (SE) aims to restore clean speech from noisy and reverberant signals~\cite{se}.
With the progress of deep learning, the field of SE has achieved remarkable breakthroughs, with algorithms generally categorized into time domain~\cite{tcnn, convtasnet, dprnn, tstnn} and time-frequency (T-F) domain~\cite{crn, dccrn, dpcrn, bsrnn, tf-gridnet, zipenhancer} methods.
Despite significantly surpassing traditional signal processing methods, deep learning-based approaches typically demand heavy computational resources, hindering their deployment on low-compute edge devices.
To address this issue, a recent line of research centers on ultra-lightweight SE model development, which typically adopts the T-F domain convolutional recurrent network (CRN)~\cite{crn} architecture, integrating various designs and techniques to reduce the overall computational cost~\cite{gtcrn, fspen, lisennet, ulunas, adaptivecrn}.
With only 30M to 100M multiply-accumulate operations per second (MACs/s), these approaches deliver promising performance, even rivaling models with substantially higher computational demands.

For resource-constrained edge devices, further reducing computational complexity remains a meaningful yet challenging task. 
To achieve this within the aforementioned CRN-based frameworks, a straightforward strategy is to compress the whole network (e.g., layers, channels, and spectral resolutions). 
However, our preliminary experiments reveal that naive downscaling may lead to rapid performance degradation, particularly in the low-frequency bands where noise components are insufficiently suppressed.
This observation suggests that a more refined architecture is needed to allocate modeling capacity across low- and high-frequency regions more effectively, thereby further reducing computational complexity while preserving low-frequency performance.
One feasible solution is to employ full/sub-band~\cite{fspen, fullsubnet} or multi-scale spectral processing~\cite{primek}, but these works lack explicit emphasis on low-frequency recovery.
\cite{firstcoarse} adopts two sub-networks to process the full-band and low-frequency regions sequentially.
Unfortunately, its strictly cascaded structure prevents the sub-networks from achieving mutual and synergistic feature fusion, and is prone to error accumulation. 
\cite{interaction} uses two parallel branches to model the magnitude spectrum and phase details with cross-domain interaction in multi-channel settings, but such architectural designs remain under-explored in ultra-lightweight monaural SE.

To this end, we propose \textbf{CoFi-Lite}, an ultra-lightweight SE model that operates in parallel at \textbf{Co}arse- and \textbf{Fi}ne-grained spectral scales while requiring only 12.87M MACs/s and 83.12k parameters.
Specifically, CoFi-Lite employs two dedicated encoder-decoder paths to process full-band envelopes and low-frequency details, respectively, thereby achieving synergistic SE performance.
We also introduce a novel Cross-Path Fusion (CPF) module, which is inserted at the bottlenecks of the two parallel paths, allowing efficient mutual feature interaction.
Extensive experiments confirm the superiority of these design choices.

\section{Methodology}

\begin{figure*}[t!]
\centering
    \includegraphics[width=0.80\linewidth]{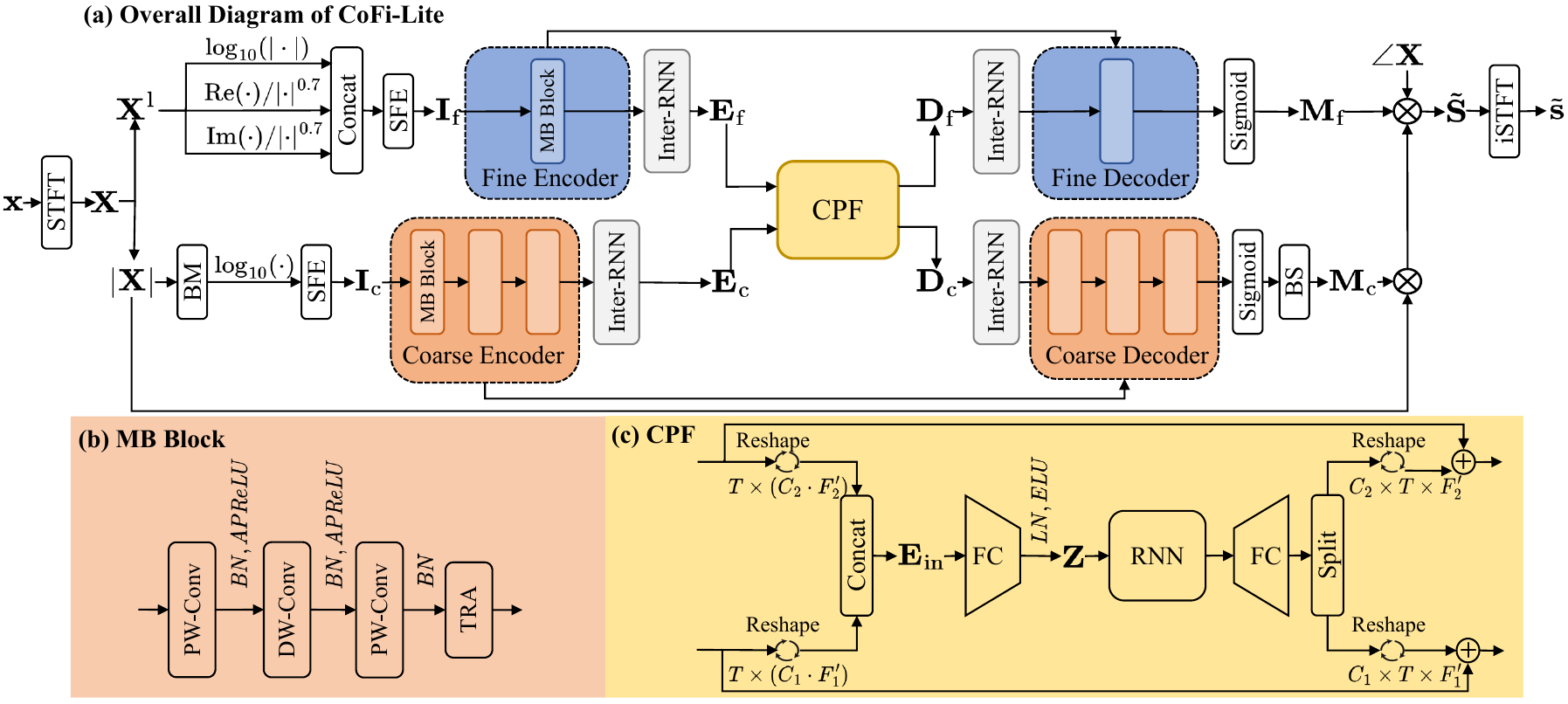}
    \caption{(a) The overall diagram of the proposed CoFi-Lite, where $\text{Re}(\cdot)$ and $\text{Im}(\cdot)$ denote the operations of extracting real and imaginary parts, respectively. (b) The details of the MB block. (c) The details of the CPF module.}
    \label{fig:cofi}
    \vspace{-1.6em}
\end{figure*}

\subsection{Model Overview}
As depicted in Fig.~\ref{fig:cofi} (a), CoFi-Lite mainly comprises two parallel coarse and fine paths, bridged by the CPF module.
Each path adopts the standard CRN framework, including an encoder, a decoder, and two inter-frame RNNs (Inter-RNNs)~\cite{dpcrn} as the bottleneck enhancers.
Given a noisy mixture $\mathbf{x}$, it is first transformed into the complex spectrum $\mathbf{X} \in \mathbb{C}^{T \times F}$ via short-time Fourier transform (STFT), where $T$, $F$ denote the time and frequency dimensions, respectively. 
$\mathbf{X}$ is then pre-processed into two distinct inputs $\mathbf{I}_{\text{c}}$ and $\mathbf{I}_{\text{f}}$ for each path. 
These features are processed by their respective encoders and the first Inter-RNN modules; the resulting representations $\mathbf{E}_{\text{c}}$ and $\mathbf{E}_{\text{f}}$ are then fed into CPF to obtain the interacted and enhanced features $\mathbf{D}_{\text{c}}$ and $\mathbf{D}_{\text{f}}$.
By mapping these output features through the remaining modules, the model predicts two ideal ratio masks (IRMs)~\cite{irm}, denoted as $\mathbf{M}_{\text{c}}$ and $\mathbf{M}_{\text{f}}$.
They are applied sequentially to recover the full-band magnitude envelope and refine low-frequency details, formulated as:
\begin{equation}
|\tilde{\mathbf{S}}(t, f)| \!=\!
\begin{cases}
|\mathbf{X}(t, f)|\otimes \mathbf{M}_{\text{c}}(t, f)\otimes \mathbf{M}_{\text{f}}(t, f), & f \!\le\! f_{\text{low}} \\
|\mathbf{X}(t, f)|\otimes \mathbf{M}_{\text{c}}(t, f), & f \!>\! f_{\text{low}}
\end{cases}
\label{eq:mag_masking}
\end{equation}
\noindent where $|\tilde{\mathbf{S}}|$ denotes the restored magnitude spectrum, and $t$, $f$ denote the frame index and frequency bin, respectively. 
The operator $\otimes$ represents element-wise multiplication, and $f_\text{low}$ is the cutoff frequency index separating low and high frequency bands. 
Notably, the target does not involve phase recovery, as ultra-lightweight models generally lack sufficient capacity for accurate phase modeling~\cite{ulunas}.
Although this choice imposes an upper bound on the theoretical performance~\cite{phase}, it ensures an effective performance-complexity trade-off.
The enhanced complex spectrum $\tilde{\mathbf{S}}$ is formed by combining $|\tilde{\mathbf{S}}|$ with the noisy phase $\angle \mathbf{X}$, followed by the inverse STFT (iSTFT) to generate the final output $\tilde{\mathbf{s}}$.
For model training, we adopt the same loss function as in~\cite{gtcrn}.

\subsection{Coarse Path}
\label{coarse}
The coarse encoder aims to capture compact features that represent coarse-grained spectral structures.
Prior to encoding, we employ the band merging (BM) module~\cite{gtcrn} to compress the sparse high-frequency information by condensing components above $f_{\text{low}}$ with an equivalent rectangular bandwidth (ERB) filter bank, formulating the coarse input $\mathbf{I}_{\text{c}}$ as:
\begin{equation}
\mathbf{I}_{\text{c}} = \mathcal{F}_{\text{SFE}} \left( \log_{10} \left( \mathcal{F}_{\text{ERB}} \left( |\mathbf{X}| \right) \right) \right)
\label{eq:coarse_input}
\end{equation}
\noindent where $\mathcal{F}_{\text{ERB}}(\cdot)$ denotes the BM operation, and $\mathcal{F}_{\text{SFE}}(\cdot)$ represents the subband feature extraction (SFE) module from~\cite{gtcrn} to boost efficient spectral utilization.
Within the encoder, three MB blocks shown in Fig.~\ref{fig:cofi} (b) are stacked to progressively reduce the frequency resolution by a factor of 2. 
Derived from~\cite{ulunas} for its proven efficacy, the MB block consists of a sequence of a point-wise convolution (PW-Conv), a depth-wise convolution (DW-Conv), and another PW-Conv, integrated with a temporal recurrent attention (TRA) module~\cite{gtcrn}. 
Notably, we replace the original causal time-frequency attention (cTFA) module with TRA for simplicity and computational efficiency. 
The configurations for Batch Normalization (BN) and affine PReLU (APReLU) are kept consistent with the original design.
The coarse decoder adopts an architecture symmetric to the encoder, utilizing transposed convolution in each DW-Conv.
Skip connections are employed between corresponding MB blocks of the encoder and decoder.
After sigmoid activation, the output undergoes the band splitting (BS) module to generate $\mathbf{M_c}$ by reversing the BM operation.

\subsection{Fine Path}
As deep compression and restricted channel count compromise the coarse path's capacity in low-frequency modeling, the fine path is introduced to recover these fine-grained details.
Unlike the coarse path, which exploits the full-band magnitude, the fine path focuses on low-frequency bands and leverages both magnitude and phase information, with its input $\mathbf{I}_{\text{f}}$ calculated as:
\begin{equation}
\mathbf{I}_{\text{f}} = \mathcal{F}_{\text{SFE}} \left( 
\bigg[ 
  \log_{10}|\mathbf{X}^\text{l}|, 
  \frac{\mathbf{X}_{\text{r}}^\text{l}}{|\mathbf{X}^\text{l}|^{0.7}}, 
  \frac{\mathbf{X}_{\text{i}}^\text{l}}{|\mathbf{X}^\text{l}|^{0.7}} 
\bigg]
\right)
\label{eq:fine_input}
\end{equation}
\noindent where $\mathbf{X}^{\text{l}}$ refers to the low-frequency bands of $\mathbf{X}$ truncated at the cutoff index $f_{\text{low}}$.
The subscripts $\text{r}$, $\text{i}$ represent the real and imaginary parts of the complex spectrum, respectively.
Dynamic range compression with an empirical exponent set of 0.7 is applied to the input components for effective feature extraction~\cite{compress, adaptivecrn, lisennet}.
Within the encoder, only a single MB block is used for feature extraction, with a (1,2) stride to downsample along the frequency dimension.
This design choice preserves the high resolution essential to low-frequency details while keeping the computational burden of subsequent modules manageable.
Analogous to the coarse path, the fine decoder mirrors the encoder's structure, and the final output utilizes a sigmoid activation to yield $\mathbf{M}_{\text{f}}$.

\subsection{Cross-Path Fusion}
\label{cpf}
To facilitate cross-path feature interaction, we introduce the CPF module.
As shown in Fig.~\ref{fig:cofi} (c), the preceding representations $\mathbf{E}_{\text{c}}$ and $\mathbf{E}_{\text{f}}$ are reshaped from $C_i \times T \times F^{\prime}_i$ to $T \times (C_i \cdot F^{\prime}_i)$, where $C_i$ and $F^{\prime}_i$ denote the channel and compressed frequency dimensions, respectively, and subscript $i \in \{1, 2\}$ corresponds to the coarse and fine paths. 
The flattened features are then concatenated to form a unified high-dimensional representation $\mathbf{E}_{\text{in}} \in \mathbb{R}^{T \times D}$, where $D = (C_{\text{1}} \cdot F'_{\text{1}}) + (C_{\text{2}} \cdot F'_{\text{2}})$.
To control the computational complexity, the CPF module first compresses $\mathbf{E}_{\text{in}}$ into an $H$-dimensional latent space using a fully connected (FC) layer.
The resulting feature is then processed by layer normalization (LN) and an exponential linear unit (ELU), after which an RNN module captures the temporal patterns of the fused feature $\mathbf{Z}$. Another FC layer is subsequently applied to restore the feature dimension to the original size $D$. 
The final output is split into two parts and reshaped back to $C_i \times T \times F'_i$, $i \in \{\text{1, 2}\}$.
Additionally, skip connections are incorporated to retain previous spectral information before obtaining $\mathbf{D}_{\text{c}}$ and $\mathbf{D}_{\text{f}}$ for each path.

\section{Experimental Setup}

\subsection{Dataset}
We train and evaluate our proposed model using the DNS3 dataset~\cite{dns3}, with the Mandarin corpus from DiDiSpeech~\cite{didispeech} as additional speech material. 
Noisy mixtures are generated by convolving clean speech with a random room impulse response (RIR) and adding a randomly selected noise clip with the SNR uniformly sampled from [-5, 15] dB.
The training target is the speech signal containing early reverberation within the first 100 ms.
For training, 72,000 noisy-clean pairs of 10-second duration (amounting to 200 hours) are generated, while 1,000 pairs are generated for validation and testing, respectively.

To further validate the generalization capability of the proposed model in different acoustic environments, we conduct additional evaluations on the official DNS Challenge 2020 test set~\cite{dns1}, covering both non-reverberant and reverberant scenarios.  
All utterances are sampled at 16 kHz.

\vspace{-0.2em}

\begin{table*}[t!]
\centering
\caption{Performance comparison on the simulated DNS3 test set. Only UL-UNAS uses the statistics reported in its original paper.}
\setlength{\tabcolsep}{1.mm}
\resizebox{0.80\linewidth}{!}{
\centering
\begin{tabular}{ccccccccccc}
    \toprule
    % 表头第一行：前6列由于跨行这里留空并在第二行填写，最后3列合并显示DNSMOS
    \multirow{2}{*}{\textbf{Models}} & \multirow{2}{*}{\textbf{Params (k)}} & \multirow{2}{*}{\textbf{MACs/s (M)}} & \multirow{2}{*}{\textbf{RTF}} & \multirow{2}{*}{\textbf{PESQ}} & \multirow{2}{*}{\textbf{ESTOI} ($\times 100$)} & \multirow{2}{*}{\textbf{SI-SNR}} & \multirow{2}{*}{\textbf{DNSMOS P.808}} & \multicolumn{3}{c}{\textbf{DNSMOS P.835}} \\
    \cmidrule(lr){9-11} % 仅在第7到第9列下方画一条短横线
    
     &  &  &  &  &  &  &  & \textbf{OVRL} & \textbf{SIG} & \textbf{BAK} \\
    \midrule

    Noisy & - & - & - & 1.40 & 66.90 & 5.61 & 2.82 & 1.63 & 2.05 & 1.86 \\
    
    \midrule 
    
    \multicolumn{10}{l}{\textbf{Level I: Below 20M MACs/s}} \\ 

    \midrule 
    
    GTCRN (Small) & \textbf{7.91} & 13.63 & 0.040 & 1.88 & 72.18 & 10.07 & 3.38 & 2.53 & 2.88 & 3.76 \\
    LiSenNet (Small) & 12.46 & 15.57 & \textbf{0.031} & 1.94 & 72.74 & 10.49 & 3.36 & 2.58 & 2.93 & 3.80 \\
    UL-UNAS (Small) & 56.43 & 13.63 & 0.054 & 2.05 & 74.87 & 11.16 & 3.47 & 2.60 & 2.94 & 3.81\\
    AdaptCRN (Small) & 34.98 & 12.97 & 0.047 & 2.06 & 75.15 & 11.19 & 3.50 & 2.65 & 2.99 & \textbf{3.85} \\
    CoFi-Lite & 83.12 & \textbf{12.87} & 0.033 & \textbf{2.16} & \textbf{76.10} & \textbf{11.80} & \textbf{3.53} & \textbf{2.70} & \textbf{3.05} & \textbf{3.85} \\
    
    \midrule
    
    \multicolumn{10}{l}{\textbf{Level II: Over 30M MACs/s}} \\
    \midrule

    GTCRN & \textbf{23.67} & \textbf{31.97} & 0.050 & 2.07 & 75.11 & 11.30 & 3.48 & 2.63 & 2.98 & 3.81 \\
    LiSenNet & 36.78 & 55.77 & \textbf{0.035} & 2.17 & 76.19 & 11.74 & 3.53 & 2.69 & 3.03 & 3.85 \\
    UL-UNAS & 171.33 & 34.91 & 0.066 & 2.25 & 77.69 & 12.07 & 3.55 & 2.69 & 3.01 & 3.86\\
    AdaptCRN & 134.51 & 40.80 & 0.053 & \textbf{2.30} & \textbf{78.15} & 12.35 & \textbf{3.59} & \textbf{2.75} & 3.08 & \textbf{3.88} \\
    CoFi-Lite (Large) & 221.31 & 32.91 & 0.036 & \textbf{2.30} & 77.94 & \textbf{12.43} & 3.56 & \textbf{2.75} & \textbf{3.09} & \textbf{3.88} \\
    
    \bottomrule
\label{tab:dns3}
\end{tabular}
}
\vspace{-1.8em}
\end{table*}

\begin{table}[t!]

\centering
\caption{Performance comparison on the DNS Challenge 2020 test set. We retrain all baselines and their variants to obtain the statistics.}
\resizebox{0.95\linewidth}{!}{
\centering
\begin{tabular}{ccccccc}
\toprule
\multirow{2}{*}{\textbf{Models}} & \multicolumn{3}{c}{\textbf{No Reverb}} & \multicolumn{3}{c}{\textbf{With Reverb}}\\
\cmidrule(lr){2-4} 
\cmidrule(lr){5-7} % 仅在第7到第9列下方画一条短横线
& \textbf{OVRL} & \textbf{SIG} & \textbf{BAK} & \textbf{OVRL} & \textbf{SIG} & \textbf{BAK} \\

\midrule

Noisy & 2.48 & 3.39 & 2.62 & 1.39 & 1.76 & 1.50\\

\midrule

\multicolumn{7}{l}{\textbf{Level I: Below 20M MACs/s}} \\ 
\midrule

GTCRN (Small) & 2.99 & 3.31 & 3.91 & 2.33 & 2.72 & 3.55\\
LiSenNet (Small) & 3.00 & 3.32 & 3.95 & 2.35 & 2.76 & 3.49\\
UL-UNAS (Small) & 3.08 & 3.36 & 4.01 & 2.39 & 2.79 & 3.52 \\
AdaptCRN (Small) & 3.11 & 3.39 & 4.02 & 2.46 & 2.85 & \textbf{3.63} \\
CoFi-Lite & \textbf{3.15} & \textbf{3.43} & \textbf{4.03} & \textbf{2.48} & \textbf{2.89} & 3.57\\

\midrule

\multicolumn{7}{l}{\textbf{Level II: Over 30M MACs/s}} \\
\midrule

GTCRN & 3.09 & 3.38 & 4.01 & 2.43 & 2.86 & 3.52\\
LiSenNet & 3.09 & 3.37 & 4.02 & 2.47 & 2.89 & 3.55\\
UL-UNAS & 3.13 & 3.40 & 4.05 & 2.44 & 2.84 & 3.56\\
AdaptCRN & \textbf{3.20} & \textbf{3.45} & \textbf{4.10} & \textbf{2.51} & \textbf{2.92} & \textbf{3.57}\\
CoFi-Lite (Large) & 3.19 & 3.44 & 4.09 & 2.50 & \textbf{2.92} & \textbf{3.57}\\

\bottomrule
\label{tab:dns1}
\end{tabular}
}
\vspace{-2.2em}
\end{table}

\subsection{Implementation Details}
\subsubsection{Parameter configuration}
The STFT is computed using a 32 ms square-root Hanning window with a hop length of 16 ms and an FFT size of 512.
The kernel size of both SFE modules is set to 3.
$f_{\text{low}}$ is set to 65, corresponding to a physical frequency of 2 kHz.
The BM module preserves the first 65 frequency bands unaltered and compresses the 192 high-frequency bands to 64 bands.
With subsequent MB blocks introducing a compression factor of 8, the coarse path reaches a total full-band compression ratio of 16.
In the coarse encoder, the first MB block uses a (3,5) kernel, while the remaining two use (1,5).
The fine encoder employs a (3,3) kernel for finer extraction.
All MB blocks have 6 output channels, and all Inter-RNN modules utilize GRUs.
CPF employs a grouped GRU (2 groups) with a latent dimension $H$ of 76.
We also investigate a scaled-up variant, CoFi-Lite~(Large), where $H$ is increased to 102, and the output channels of the coarse and fine encoders are expanded to [6, 12, 14] and 14, respectively.
All other settings remain unchanged.

\subsubsection{Training configuration}
The models are trained using the Adam optimizer~\cite{adam} with a linear warmup scheduler followed by cosine annealing.
The training procedures last for 200 epochs, where each epoch contains 1,250 iterations with a batch size of 8.
The learning rate increases linearly from $10^{-6}$ to $10^{-3}$ over the initial 25,000 iterations and then decays following a cosine schedule until 250,000 iterations.  

\vspace{-0.3em}

\subsection{Baseline Models and Evaluation Metrics}
We select the latest ultra-lightweight SE models as our baselines, including GTCRN~\cite{gtcrn}, LiSenNet~\cite{lisennet}, UL-UNAS~\cite{ulunas}, and AdaptCRN~\cite{adaptivecrn}.
In addition, we include their scaled-down variants to provide a fairer comparison.
Where available, we use reported statistics from the original papers.
For missing results, we retrain the models using original codebases and parameter configurations, whereas for scaled-down variants, we proportionally reduce the number of channels for model compression. 
All retrained models use our own training configuration—except for AdaptCRN and its variant, which follow the settings from the original work.

Considering the real-time requirements of edge devices, we measure the real-time factor (RTF) using ONNX Runtime~\cite{onnxruntime} on an Intel(R) Core(TM) i5-14600KF with streaming inference. STFT and iSTFT operations are excluded, and no future frames are used, ensuring zero algorithmic delay.
To evaluate the SE performance, three intrusive metrics are employed, including scale-invariant signal-to-noise ratio (SI-SNR)~\cite{sisnr}, wide-band perceptual evaluation of speech quality (PESQ)~\cite{pesq} and extended short-time objective intelligibility (ESTOI)~\cite{estoi}. 
Additionally, two non-intrusive metrics, DNSMOS P.808~\cite{p808} and DNSMOS P.835~\cite{p835}, are utilized to assess the enhanced speech quality. DNSMOS P.835 includes three sub-metrics: OVRL for overall speech quality, SIG for signal quality, and BAK for background noise quality.

\section{Experimental Results}

\subsection{Comparison With the Baseline Models}
The results on the simulated DNS3 test set are summarized in Table~\ref{tab:dns3}, categorized by complexity levels\footnote{The parameters and MACs/s are measured by the \texttt{ptflops} toolkit: \url{https://github.com/sovrasov/flops-counter.pytorch}, with a detailed per-module breakdown available on our aforementioned demo page.} (Level~I/II).
In Level~I, CoFi-Lite demonstrates a distinct advantage over the scaled-down variants of baselines.
It also clearly outperforms the Level II baseline GTCRN in PESQ (+0.09), OVRL (+0.07), and SIG (+0.07), while requiring only 40.26\% of its computational cost and reducing RTF by 34.00\%.
In Level~II, CoFi-Lite (Large) maintains competitive efficacy against AdaptCRN with a 19.34\% reduction in computational demands, while also recording one of the lowest RTFs.
Despite exhibiting a higher parameter count, it remains within an acceptable range for deployment on most edge devices~\cite{tgru}.

Results on the official DNS 2020 test set are shown in Table~\ref{tab:dns1}. Notably, intrusive metrics are excluded due to a mismatch between the anechoic reference and our training objective, which includes early reverberation.
As shown, the performance trends align closely with the results observed on the DNS3 dataset.

\subsection{Ablation Study}
\subsubsection{Effect of key design choices}
In this section, we investigate two key designs of our model: (i) the architecture of the two parallel paths, and (ii) the CPF module.
The results are presented in Table~\ref{ablation} (IDs~1--5).
For a fair comparison, we align the complexity across all models by adjusting the hidden size of the Inter-RNNs.
Specifically, ID4 scales up ID3 to match the parameters of our proposed ID5.
As shown, ID3 outperforms single-path models (IDs~1-2).
Crucially, ID5 achieves a substantial PESQ gain (+0.14) over ID3 by integrating CPF.
Ruling out the impact of increased parameters, it still yields a PESQ advantage (+0.10) over ID4.
This indicates that $\mathbf{E}_{\text{c}}$ and $\mathbf{E}_{\text{f}}$ are highly complementary, and their interaction creates a strong synergy that boosts model performance.

\subsubsection{Investigation of different compression settings}
Table~\ref{ablation} (IDs~6--9) details the performance across different spectral compression ratios.
Let $\mathbf{R}_{\text{c}}$ and $\mathbf{R}_{\text{f}}$ denote the compression ratios of the coarse and fine paths, respectively.
Based on ID5, $\mathbf{R}_{\text{c}}$ and $\mathbf{R}_{\text{f}}$ are adjusted by adding or removing MB blocks with a (1,5) kernel and a (1,2) stride in the corresponding encoder and decoder.
It can be observed that the model performance rapidly degrades as $\mathbf{R}_{\text{f}}$ increases (IDs~5--7).
We attribute this degradation to the severe information loss caused by deep compression, which is unsuitable for the fine path designed to model low-frequency details.
Conversely, increasing the frequency resolution in the coarse path results in marginal or negligible improvements (IDs~8--9 vs. ID5).
This aligns with intuition, as envelope enhancement does not require capturing excessive fine-grained spectral details.
\subsubsection{Study of various cutoff frequency settings}
The impact of varying $f_{\text{low}}$ settings on performance is explored in Table~\ref{ablation} (IDs~10--12).
$\mathbf{R}_{\text{c}}$ remains unchanged by adjusting the compression ratio of the BM module.
Notably, increasing $f_{\text{low}}$ from 17 to 65 consistently yields performance gains, as it allows the fine path to capture more low-frequency details.
However, no additional gains are obtained when increasing $f_{\text{low}}$ to 97, since salient speech structures are concentrated below 2 kHz, leaving higher bands with sparse informative content.

\begin{table}[t!]
\centering
    \caption{Ablation study results of the coarse/fine path design, the CPF module, different compression ratio and cutoff frequency settings. ``$-$'' indicates that the corresponding path is removed.}
\setlength{\tabcolsep}{0.8mm}
\resizebox{\linewidth}{!}{
    \begin{tabular}{cccccccccc}
        \toprule
        \textbf{IDs} & $\mathbf{R}_{\text{c}}$ & $\mathbf{R}_{\text{f}}$ & \textbf{$f_{\text{low}}$} & \textbf{CPF} & \textbf{Params (k)} & \textbf{MACs/s (M)} & \textbf{PESQ} & \textbf{ESTOI} ($\times 100$) & \textbf{SI-SNR} \\
        \midrule
        1 & 16 & - & 65 & \ding{55} & 19.71 & 12.37 & 1.97 & 73.73 & 10.68 \\
        2 & - & 2 & 65 & \ding{55} & 9.24 & 12.05 & 1.53 & 70.14 & 8.41 \\
        3 & 16 & 2 & 65 & \ding{55} & 21.23 & 12.24 & 2.02 & 74.06 & 10.81 \\
        4 & 16 & 2 & 65 & \ding{55} & 90.80 & 13.44 & 2.06 & 74.98 & 11.20 \\
        5 & 16 & 2 & 65 & \ding{51} & 83.12 & 12.87 & 2.16 & 76.10 & 11.80 \\
        \midrule
        6 & 16 & 4 & 65 & \ding{51} & 71.18 & 11.89 & 2.12 & 75.72 & 11.55 \\
        7 & 16 & 8 & 65 & \ding{51} & 66.20 & 11.46 & 2.09 & 75.31 & 11.41 \\
        \midrule
        8 & 8 & 2 & 65 & \ding{51} & 95.06 & 13.84 & 2.16 & 76.10 & 11.73 \\
        9 & 32 & 2 & 65 & \ding{51} & 78.14 & 12.44 & 2.14 & 76.01 & 11.77 \\
        \midrule
        10 & 16 & 2 & 17 & \ding{51} & 58.79 & 11.51 & 2.08 & 75.30 & 11.39 \\
        11 & 16 & 2 & 33 & \ding{51} & 66.90 & 11.90 & 2.11 & 75.69 & 11.58 \\
        12 & 16 & 2 & 97 & \ding{51} & 99.35 & 14.09 & 2.15 & 76.25 & 11.72 \\
        \bottomrule
    \end{tabular}
}
\label{ablation}
\vspace{-1.2em}
\end{table}

\section{Conclusion}
In this letter, we present CoFi-Lite, a highly efficient SE model requiring extremely low computational resources.
The architecture effectively decouples full-band envelopes and low-frequency details via parallel and symmetric paths for complementary enhancement.
In addition, a novel CPF module is introduced to facilitate mutual feature interaction.
Remarkably, CoFi-Lite outperforms GTCRN with significantly reduced computational complexity. 
Its scaled-up variant also achieves competitive performance among existing ultra-lightweight SE models, further confirming the potential of our proposed method. 
However, current measurements (e.g., RTF on a desktop CPU) might not directly reflect real-world deployment on platforms such as ARM Cortex-M and DSP, and closing this gap is our future work.

\bibliographystyle{IEEEtran}
\bibliography{ref}

\end{document}